\documentclass[conference]{IEEEtran}

\usepackage{cite}
\usepackage{amsmath,amssymb}
\usepackage{graphicx}
\usepackage{booktabs}
\usepackage{multirow}
\usepackage{microtype}

\title{The Algorithmic Barrier: A Framework for Artificial\\ Frictional
Unemployment and Information Asymmetry in\\ Automated Recruitment Systems}

\author{
\IEEEauthorblockN{Ibrahim Denis Fofanah}
\IEEEauthorblockA{
Seidenberg School of Computer Science and Information Systems\\
Pace University\\
New York, USA\\
Email: if57774n@pace.edu
}
}

\begin{document}
\maketitle

\begin{abstract}
The United States labor market has entered a period in which high job vacancy rates and prolonged unemployment persist together. Classical theory attributes such conditions to skills mismatch or geographic immobility, but neither fully explains a pattern now widely reported: qualified candidates are rejected at the earliest, automated stage of hiring, before any human sees their application. This paper introduces \emph{Artificial Frictional Unemployment} (AFU), a framework describing how deterministic automated screening rejects qualified candidates through semantic misinterpretation rather than genuine skill gaps. We situate the phenomenon within labor economics and information asymmetry theory and formalize the mechanism by which legacy Applicant Tracking Systems (ATS) turn hiring into a high-precision classification problem that inflates false negatives.

The contribution is primarily conceptual. To make the mechanism concrete, we report a controlled proof-of-concept simulation comparing keyword-based screening with vector-space semantic matching under identical conditions. The simulation shows how lexical variance alone can produce false negatives; it is not a measurement of how much real-world friction is artificial, which we leave to future field studies. Building on the framework, we outline JobOS, a candidate-side architecture that illustrates how semantic competency mapping could operate alongside existing hiring infrastructure. Framing automated recruitment as labor market infrastructure, rather than a firm-level convenience, exposes a correctable source of matching inefficiency with consequences for workforce participation and the use of human capital.
\end{abstract}

\begin{IEEEkeywords}
Artificial Frictional Unemployment, Applicant Tracking Systems, Automated Hiring, Semantic Matching, Labor Market Efficiency, Information Asymmetry, Vector Embeddings
\end{IEEEkeywords}

\section{Introduction}
The United States labor market has displayed a persistent puzzle. Employers report difficulty filling positions while many workers remain unemployed or underemployed for long stretches. Under conventional theory this should not last: when vacancies are plentiful, workers and firms match quickly. The mechanism no longer behaves that way.

The Beveridge Curve, which relates unemployment to vacancies, offers a useful lens. Movements along it track the business cycle; outward shifts signal declining matching efficiency. Since the early 2020s the curve has shifted outward, with vacancies and unemployment coexisting at levels standard models struggle to reconcile. The difficulty appears to lie less in the supply of or demand for labor than in the process that pairs the two.

Skills mismatch, geographic immobility, demographic change, and wage rigidity remain part of the explanation. They do not account for candidates who meet stated requirements yet are rejected before any human review. When rejection precedes human judgment, the matching process itself becomes a candidate for scrutiny.

We argue that a meaningful share of contemporary frictional unemployment is artificial: produced by the design of automated recruitment rather than by any real incompatibility between worker and firm. ATS have become central to hiring, especially in large organizations, where they were adopted to manage scale and limit risk. In doing so they moved the decisive filtering step upstream, ahead of contextual judgment.

Most legacy platforms screen deterministically---exact keyword matches, rigid rules, fixed thresholds. Engineered to minimize false positives, they carry a structural bias toward over-rejection, because rejecting a qualified applicant is treated as far cheaper than advancing an unqualified one. The consequence is a steady stream of false negatives that no one observes, since rejected candidates rarely learn why.

The economic cost follows directly. Candidates filtered out for unusual terminology, non-linear careers, or employment gaps are not mismatched in the classical sense. They are misinterpreted, and repeated misinterpretation lengthens searches, dampens mobility, and leaves capable labor idle. Information economics sharpens the point: hiring is a market of severe information asymmetry, and rigid filters, adopted to manage that uncertainty, tend to deepen it. As screening hardens, resumes lose value as signals, which invites still stricter filtering.

At the center sits a semantic gap between how people describe skills and how systems read them. Competencies are rarely single tokens. Equivalent abilities are phrased differently across industries, backgrounds, and cultures, and deterministic systems read that variation as absence rather than equivalence. Two candidates with the same qualifications can be sorted differently on wording alone.

Closing this gap means treating hiring less as filtering and more as translation. On that view we outline JobOS, a candidate-side architecture that standardizes, verifies, and semantically aligns competencies with employer intent. We present it as a design proposal illustrating how the framework could be built, not as a system evaluated here.

This work makes three contributions. First, it introduces and formalizes Artificial Frictional Unemployment, placing semantic misinterpretation within established labor market theory as a digitally induced matching failure. Second, it reports a controlled proof-of-concept simulation that isolates the mechanism by which vocabulary variance produces false negatives under deterministic screening, and how semantic representation mitigates it. Third, it outlines JobOS as a design proposal for operationalizing semantic competency mapping alongside existing infrastructure.

One scope note governs the rest of the paper. The simulation isolates a mechanism under controlled, synthetic conditions; its numbers demonstrate behavior, not the prevalence of AFU in the economy. Estimating that prevalence requires field data and is left to future work.

\section{Literature Review}
Three bodies of work meet in this study: labor market research on underutilized workers, economic theory on information asymmetry, and computational analysis of algorithmic hiring. Read together, they describe a recurring failure---hiring technologies built for efficiency operate on assumptions that clash with how skills are actually expressed.

\subsection{The ``Hidden Worker'' Phenomenon}
A large population of workers is willing and able to work yet remains locked out of opportunities. The Harvard Business School study by Fuller et al.~\cite{fuller2021} counts more than 27 million such ``hidden workers'' in the United States, screened out not for lack of ability but by automated processes. The study attributes much of this to rigid filters: demands for continuous employment, narrow credential definitions, and exact keyword matches. Candidates with non-linear paths, caregiving gaps, or military service are hit hardest, even when their skills fit. Exclusion of this kind is a failure of skill \emph{recognition}, not skill supply---automated filters work on impoverished representations that miss the equivalence between different expressions of the same competency.

\subsection{Information Asymmetry and Labor Market Signaling}
Akerlof's analysis of the market for lemons~\cite{akerlof1970} shows how asymmetric information pushes markets toward inefficiency: unable to tell quality apart, buyers discount everything, and good offerings withdraw. Hiring behaves similarly. Employers cannot verify claims before interviews, and candidates signal through resumes that are self-reported and inconsistent. Automated screening is the response to that uncertainty, but by enforcing narrow definitions of qualification it reproduces Akerlof's outcome, defaulting to conservative thresholds that exclude many qualified applicants.

\subsection{Algorithmic Bias and Deterministic Screening}
Work in computer science and policy has scrutinized how these systems behave. Bogen and Rieke~\cite{bogen2018} document the reliance of many ATS on Boolean matching and rule-based scoring, which assume that skills reduce cleanly to fixed terms. They do not. Equivalent competencies vary by industry and background, and deterministic systems read that variation as a substantive difference, producing systematic misclassification. Fairness research~\cite{raghavan2020,sanchezmonedero2020} shows the burden falls disproportionately on candidates from non-traditional backgrounds, where vocabulary rather than skill drives the outcome. Regulation has begun to respond, including New York City's Local Law 144 on automated employment decision tools~\cite{nyc2021ll144} and the European Union's AI Act, which treats employment-related AI as high-risk~\cite{euaiact2024}.

\subsection{Semantic Matching Architectures}
NLP has long worked to move past keyword matching. Early statistical methods---Latent Semantic Analysis, Latent Dirichlet Allocation---captured co-occurrence but missed the fine distinctions hiring demands, such as ``Java'' the language versus ``Java'' the place. Distributed representations, Word2Vec~\cite{mikolov2013} and GloVe~\cite{pennington2014}, measured similarity in dense vector spaces but assigned one vector per term and so could not resolve polysemy. Transformer models, notably BERT~\cite{devlin2019}, produce contextual embeddings through self-attention~\cite{vaswani2017}, and sentence encoders such as Sentence-BERT~\cite{reimers2019} extend this to efficient document-level comparison. The remaining tension is practical: full contextual inference is costly at recruitment scale, forcing a choice between speed and nuance.

\subsection{Research Gap}
Each field sees part of the problem. Labor economics counts the excluded; information theory explains the incentive to over-filter; computer science shows how deterministic screening enacts it and why accurate semantic models are hard to deploy cheaply. Recent work has documented the systemic labor-market costs of ATS, including language mismatch and prolonged unemployment~\cite{trumble2025}. What is missing is a treatment that names this exclusion as a distinct category of frictional unemployment and frames the latency--accuracy tradeoff as a question of labor market infrastructure. Our contribution is therefore not the observation that automated screening excludes qualified candidates, but the framing of that exclusion as Artificial Frictional Unemployment---a nameable inefficiency situated within labor-market theory and synthesized with information-asymmetry economics and the semantic-matching mechanism. This paper bridges those perspectives.

\section{Methodology}
We model candidate screening as a classification problem and isolate semantic interpretation as the variable of interest. The section formalizes legacy ATS as deterministic classifiers, describes a semantic representation based on vector embeddings, and specifies the controlled simulation that compares them.

The purpose of the simulation is narrow and worth stating plainly: to isolate one mechanism---how lexical variance alone produces false negatives under exact-match screening---under fully controlled conditions. Because the data are built to contain semantic variation, the direction of the result is expected. The value lies in making the mechanism explicit, reproducible, and open to inspection, not in estimating its economic size.

\subsection{ATS as a Deterministic Classifier}
Legacy ATS act as early-stage classifiers with a binary output: advance or reject. Their logic rests on Boolean matching, weighted term frequencies, and rule-based thresholds. This favors precision over recall---strict matching suppresses false positives at the cost of a higher false negative rate---and the resulting errors go unrecorded, since rejected candidates receive no feedback. Formally, we treat screening as
\[
f_{\text{ATS}}: C \rightarrow \{0,1\},
\]
where \(C\) is a candidate's resume and \(1\) denotes advancement. Semantic equivalence between different expressions of a skill is preserved only through exact term overlap, so vocabulary variance becomes a dominant source of error.

\subsection{Semantic Representation}
A semantic approach maps resumes and job descriptions into a shared vector space, where similarity is geometric rather than lexical. With \( \mathbf{v}_C \in \mathbb{R}^n \) the resume embedding and \( \mathbf{v}_J \in \mathbb{R}^n \) the job embedding, alignment is measured by cosine similarity:
\[
S(C, J) = \frac{\mathbf{v}_C \cdot \mathbf{v}_J}{\|\mathbf{v}_C\| \, \|\mathbf{v}_J\|}.
\]
This captures relational meaning: ``statistical modeling with Python'' can align with ``machine learning using scikit-learn'' despite no shared keyword.

\subsection{Contextual Embeddings}
Contextual models refine this further by letting surrounding text shape each token's representation, computed through self-attention:
\[
\text{Attention}(Q, K, V) = \text{softmax}\!\left(\frac{QK^\top}{\sqrt{d_k}}\right)V,
\]
with \(Q\), \(K\), \(V\) the query, key, and value matrices. Context lets ``lead'' read as a verb (``led a team'') or a noun (``lead engineer''), which reduces the noise that degrades matching.

\subsection{Controlled Simulation}
We compare keyword screening with semantic matching on a synthetic dataset built to isolate semantic effects, since proprietary ATS logs are not available for controlled study. The dataset holds \(N = 1{,}000\) resume--job pairs. Crucially, ground truth is defined on skills, not words: each job carries a required skill set, each candidate a possessed skill set, and a pair is labeled ``qualified'' when the candidate covers at least 80\% of the job's required skills. Only after labeling is the text rendered, with each skill expressed through a randomly chosen surface form---synonym, acronym, or title variant. Lexical variation therefore enters the wording while the label stays tied to the underlying competencies. Job descriptions are held constant across conditions.

Two pipelines run on the same data. The keyword baseline extracts the required terms from each job description and advances a candidate when the fraction of those terms matched by whole-word comparison reaches 0.4. The semantic pipeline ranks candidates by cosine similarity and advances those above a threshold \(\tau\). We report precision, recall, and F1, with attention to false negatives.

\subsection{Reproducibility}
Pairs are drawn from a vocabulary of 35 skills, each with associated synonyms, acronyms, and title variants. The semantic pipeline uses the open sentence-transformer \texttt{all-MiniLM-L6-v2}, which produces 384-dimensional embeddings compared by cosine similarity; the main comparison uses a decision threshold of 0.49. The keyword baseline is the whole-word required-term rule described above, with an advancement threshold of 0.4. Labels follow the 80\% skill-coverage rule, independent of wording. The experiment runs from a fixed random seed, so any rerun reproduces the reported numbers and figures exactly.
The experiment code and synthetic-data generator are available from the author to support independent replication.

\subsection{Scope and Assumptions}
The design fixes everything except representation and matching logic---candidate behavior, recruiter judgment, and downstream interview performance are excluded---so that the effect of representation is visible on its own. Synthetic data limits external validity but buys precise control over linguistic variation and skill equivalence. The method suits the identification of a structural failure mode, not the estimation of its prevalence.

\section{Proposed Architecture: JobOS}
JobOS illustrates how the framework of Section~I might be operationalized. We present it as a design proposal: the architecture is described to show that semantic competency mapping can function as infrastructure, and it is not evaluated as a deployed system here. The design interoperates with existing ATS rather than replacing them, targeting their dominant failure mode.

The system is organized into four layers---ingestion and normalization, semantic representation and indexing, verification and simulation, and secure storage and governance---each able to evolve independently.

\begin{figure}[htbp]
\centering
\includegraphics[width=\linewidth]{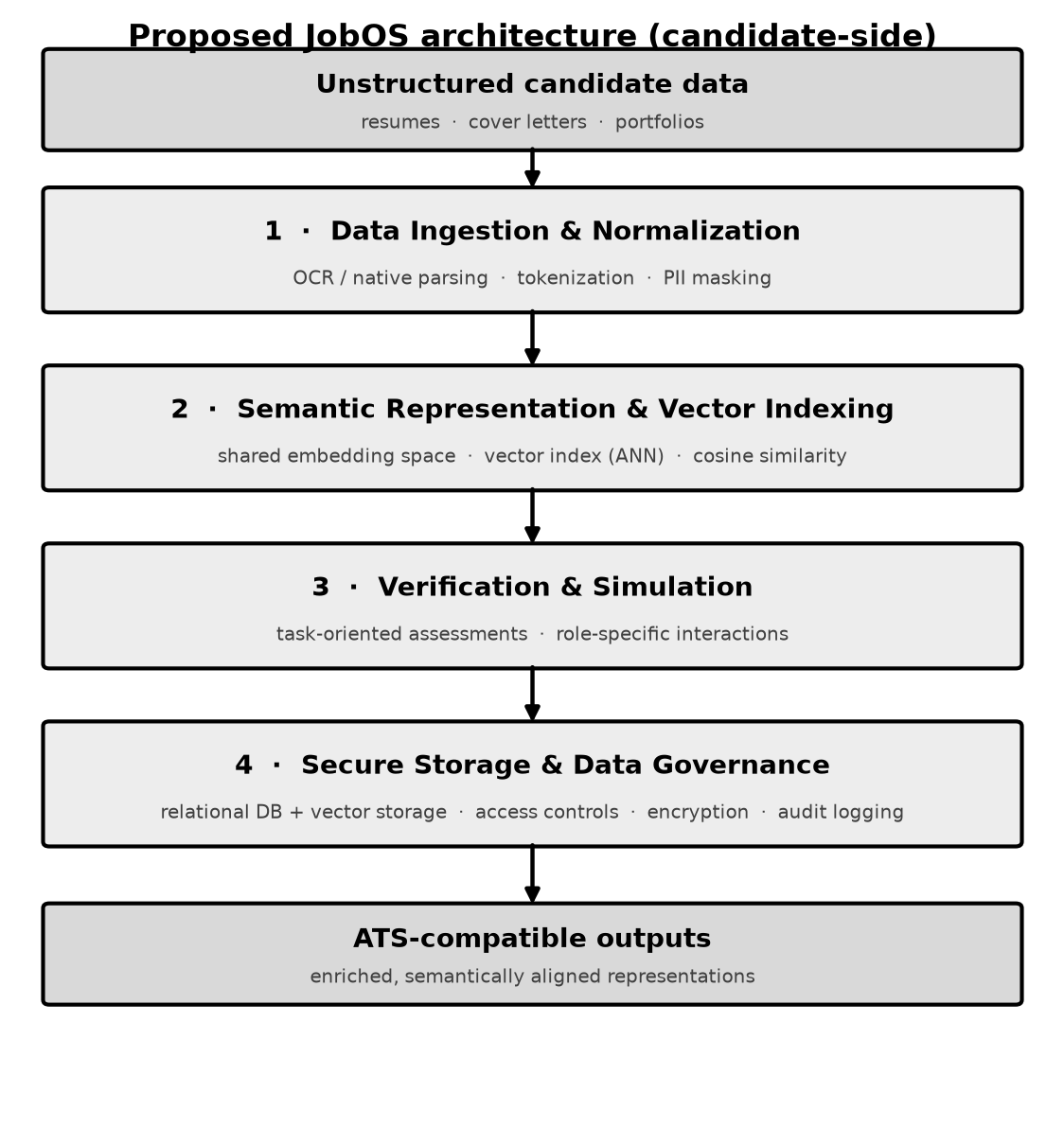}
\caption{Proposed JobOS architecture: a candidate-side pipeline from unstructured candidate data to ATS-compatible outputs. Shown as a design proposal, not an evaluated system.}
\label{fig:jobos_architecture}
\end{figure}

\subsection{Ingestion and Normalization}
The ingestion layer converts resumes, cover letters, and portfolios---typically PDF or DOCX---into a standardized representation. Text is extracted by OCR for scanned files and native parsers for digital ones, then tokenized, de-duplicated, and segmented. Personally identifiable information is masked at this stage to support blind downstream evaluation. What emerges is a structured intermediate form that keeps semantic content while discarding formatting.

\subsection{Semantic Representation and Indexing}
Normalized data is embedded into the shared vector space of Section~III and stored in an index built for approximate nearest-neighbor search, so the system retrieves a ranked shortlist by cosine similarity instead of comparing every pair. Embeddings are computed once and reused; this amortization is the design's answer to the latency--accuracy tension, keeping contextual accuracy without paying inference cost at query time.

\begin{figure}[htbp]
\centering
\includegraphics[width=\linewidth]{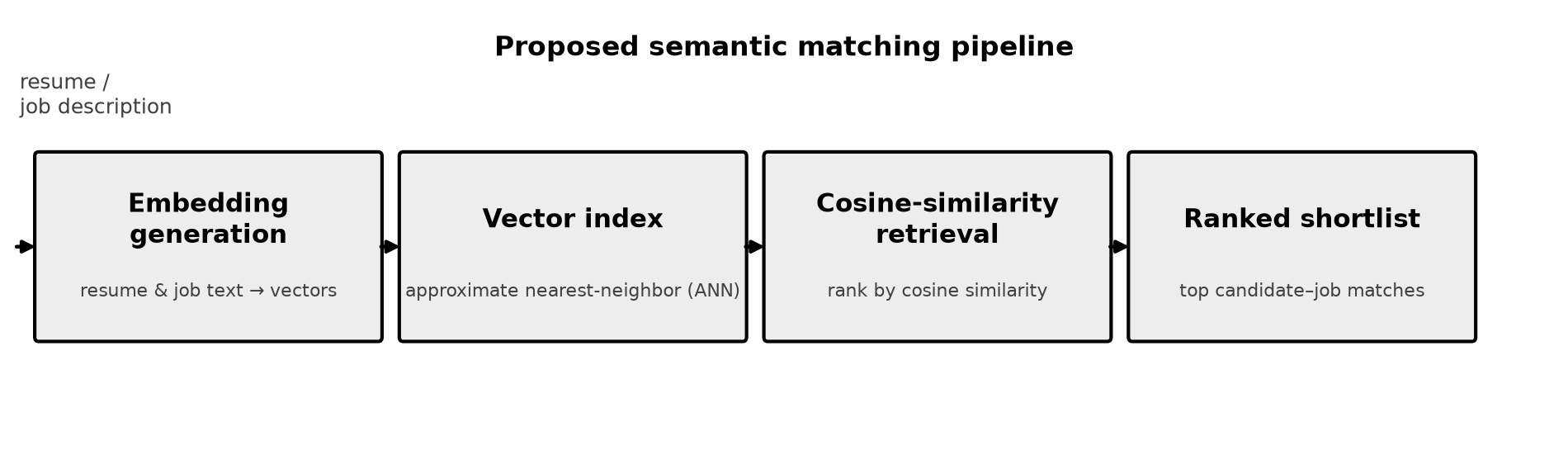}
\caption{Proposed semantic matching pipeline: embedding generation, vector indexing, and similarity-based retrieval.}
\label{fig:vector_pipeline}
\end{figure}

\subsection{Verification and Simulation}
A verification layer addresses information asymmetry by testing competencies rather than trusting self-report. As proposed, it would run structured, role-specific assessments that approximate an early-stage evaluation, capturing responses through text and, in a planned extension, speech, and treating auxiliary signals such as response latency as complementary evidence. The layer is designed to augment human judgment, not to replace it. It is described here as design; it is not part of the evaluation in this paper.

\subsection{Storage and Governance}
Employment data is sensitive, so governance is treated as a first-class concern. Candidate data, embeddings, and verification artifacts would sit in a relational store with vector support and fine-grained access control. The governing principle is candidate ownership: candidates control disclosure and can revoke access, with encryption at rest and in transit and audit logging throughout.

\begin{figure}[htbp]
\centering
\includegraphics[width=\linewidth]{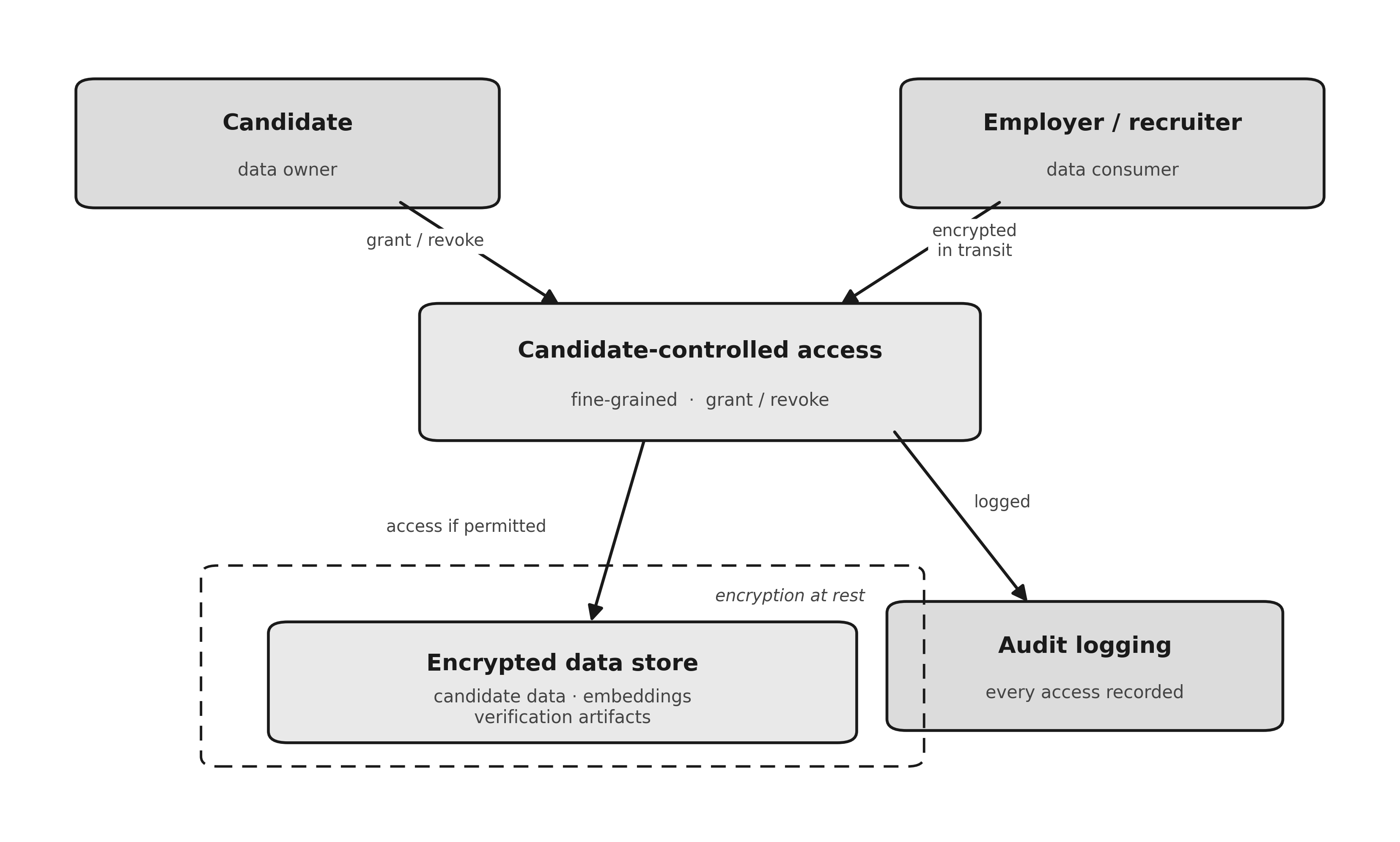}
\caption{Proposed JobOS data governance model: candidate-controlled access, encryption boundaries, and audit logging.}
\label{fig:data_governance}
\end{figure}

\subsection{Prototype Feasibility}
To check feasibility, we built an alpha prototype on a cloud-native stack: a React.js frontend, serverless orchestration over PostgreSQL, and the \texttt{pgvector} extension for vector indexing, with semantic parsing handled by the Gemini 2.0 Flash API. The prototype shows that the pipeline can be assembled from commercially available components. It is a feasibility check, not a measurement of the system's effect on hiring outcomes.

\section{Illustrative Simulation Results}
This section reports the controlled simulation of Section~III. The aim is to show how semantic matching changes screening outcomes relative to deterministic keyword screening. Every figure and table is generated directly from the run ($N = 1{,}000$, fixed seed), and the reported values characterize behavior on synthetic data built to contain semantic variation---a demonstration of the mechanism, not an estimate of real-world friction.

\subsection{Robustness to Lexical Variation}
\begin{figure}[htbp]
\centering
\includegraphics[width=\linewidth]{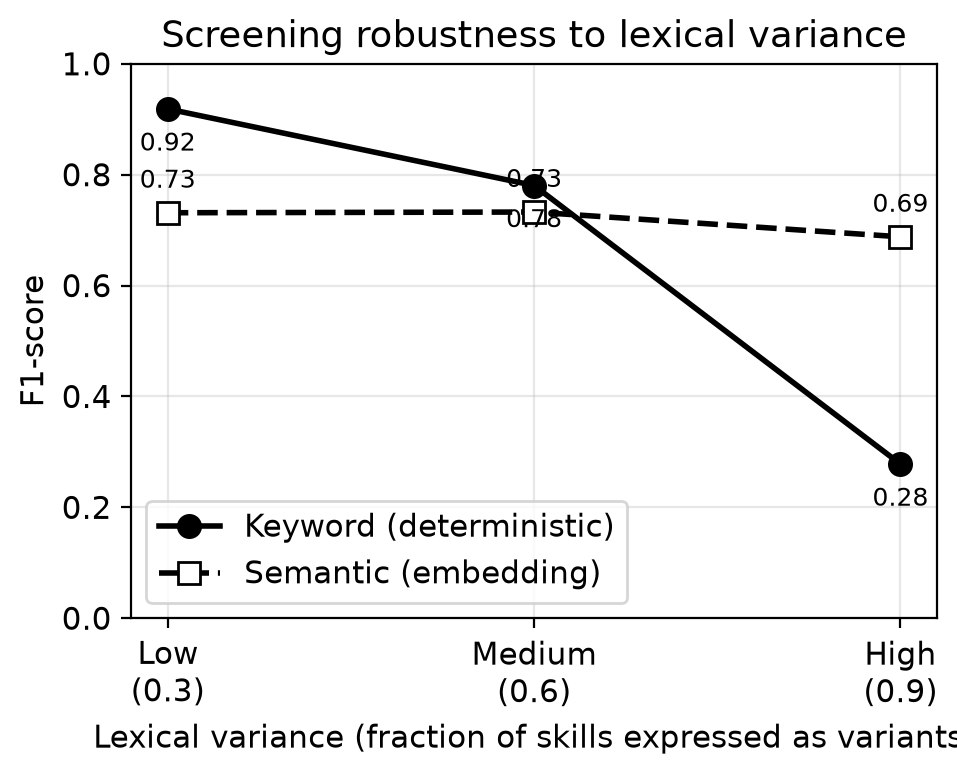}
\caption{F1 against lexical variance. Keyword screening collapses as vocabulary diversifies ($0.92 \rightarrow 0.28$); semantic matching holds ($0.73 \rightarrow 0.69$). The crossover illustrates the AFU mechanism.}
\label{fig:lexical_robustness}
\end{figure}

The mechanism is clearest when lexical variance rises while qualifications stay fixed. Figure~\ref{fig:lexical_robustness} tracks F1 for both pipelines across low, medium, and high variance. The keyword baseline is strong when candidates use standard terms (F1 $=0.92$) but falls to $0.28$ as vocabulary diversifies. The semantic pipeline barely moves, from $0.73$ to $0.69$.

That crossover is Artificial Frictional Unemployment made visible. Keyword screening cannot absorb variation in how a skill is phrased; as wording drifts from the job description, exact matching reads qualification as absence. Semantic representation is largely indifferent to surface form. The finding also sharpens the hidden-worker argument: candidates from non-traditional backgrounds are the ones most likely to use non-standard vocabulary, and so the ones this failure mode most reliably excludes.

\subsection{Comparative Performance}
At the main operating point (70\% lexical variance), the two pipelines split cleanly, as Table~\ref{tab:performance_comparison} shows. The keyword baseline is precise (0.90) but low on recall (0.52): it lets through few unqualified candidates and rejects nearly half the qualified ones. The semantic pipeline reverses the profile, reaching 0.90 recall at 0.61 precision. Its F1 (0.72) edges past the baseline's (0.66), but the more telling story is how that edge is bought.

\begin{table}[htbp]
\caption{Screening Performance ($N=1{,}000$, 70\% Lexical Variance)}
\label{tab:performance_comparison}
\centering
\begin{tabular}{lccc}
\toprule
\textbf{Metric} & \textbf{Keyword} & \textbf{Semantic} & \textbf{Difference} \\
\midrule
Precision & 0.90 & 0.61 & $-0.29$ \\
Recall    & 0.52 & 0.90 & $+0.38$ \\
F1-Score  & 0.66 & 0.72 & $+0.06$ \\
\bottomrule
\end{tabular}
\end{table}

\subsection{The Recall--Precision Tradeoff}
Semantic matching is not uniformly better. It buys a large recall gain ($0.52 \rightarrow 0.90$) by spending precision ($0.90 \rightarrow 0.61$). Of 500 qualified candidates, the keyword baseline rejected 239---a 48\% false negative rate---while the semantic pipeline cut false negatives to 51. Recovering those candidates is the point, and the cost is that semantic matching also admits more false positives.

\begin{figure}[htbp]
\centering
\includegraphics[width=\linewidth]{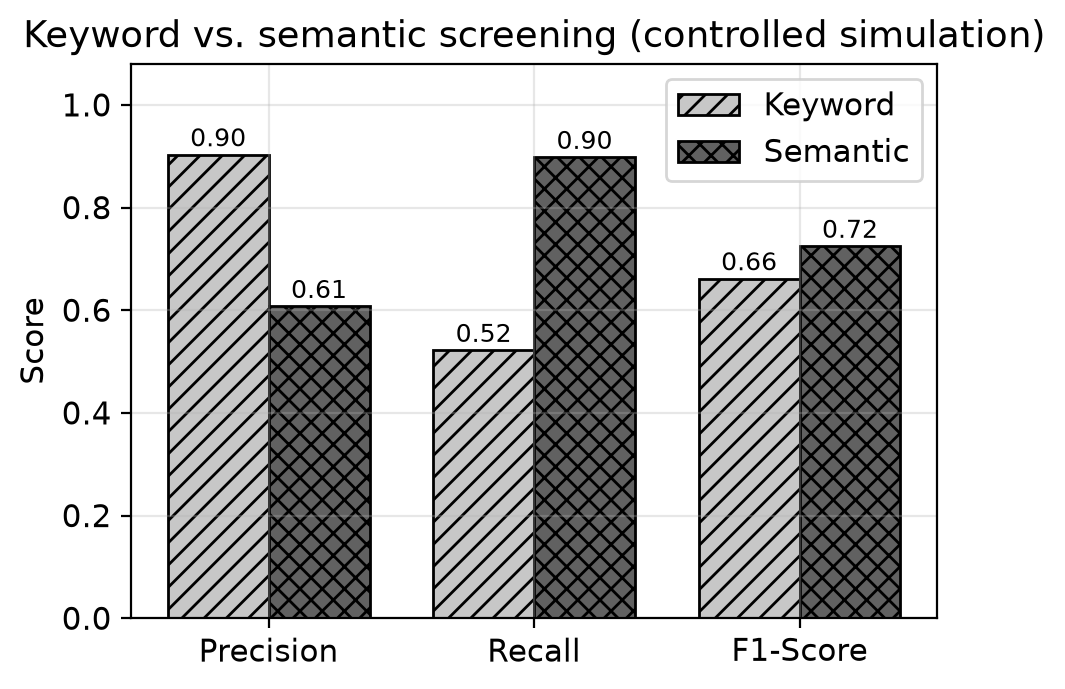}
\caption{Precision, recall, and F1 for keyword and semantic screening. Semantic matching trades precision ($0.90 \rightarrow 0.61$) for recall ($0.52 \rightarrow 0.90$).}
\label{fig:metrics_comparison}
\end{figure}

This tradeoff is a familiar property of similarity-based retrieval, and stating it plainly matters both for honesty and for design. A method that maximizes recall at some cost in precision belongs upstream of human review, surfacing candidates a person then evaluates, rather than making unattended accept-or-reject calls. That is precisely the role we propose for JobOS in Section~IV. Figure~\ref{fig:metrics_comparison} shows the full picture, precision cost included.

\subsection{Threshold Sensitivity}
\begin{figure}[t]
\centering
\includegraphics[width=\linewidth]{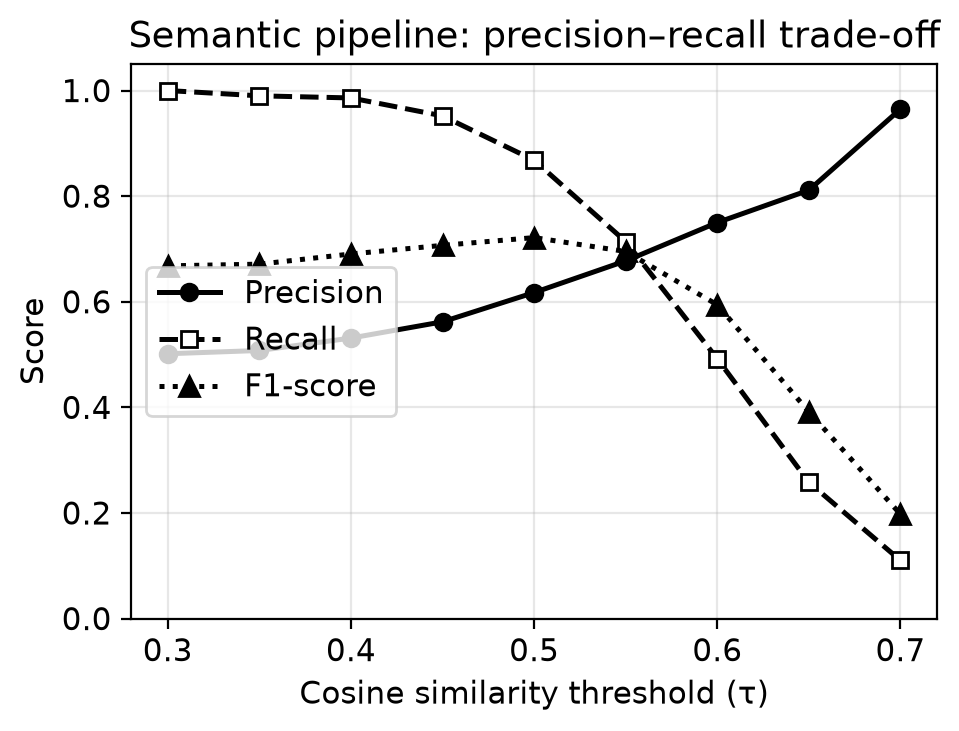}
\caption{Precision, recall, and F1 across the similarity threshold \(\tau\) for semantic matching. F1 peaks near \(\tau \approx 0.50\).}
\label{fig:threshold_sweep}
\end{figure}

The semantic threshold \(\tau\) is tunable, so the precision--recall balance can be set to context. Table~\ref{tab:threshold_sensitivity} and Figure~\ref{fig:threshold_sweep} trace precision, recall, and F1 as \(\tau\) varies from 0.30 to 0.70. Recall falls from near-total (1.00 at \(\tau = 0.30\)) to 0.11 at \(\tau = 0.70\), while precision climbs from 0.50 to 0.96; F1 peaks at 0.72 near \(\tau \approx 0.50\). This is what lets a deployment choose an operating point---high recall where the goal is to surface candidates for review, higher precision where reviewer capacity is scarce.

\begin{table}[t]
\caption{Threshold Sensitivity for Semantic Matching}
\label{tab:threshold_sensitivity}
\centering
\begin{tabular}{cccc}
\toprule
$\boldsymbol{\tau}$ & \textbf{Precision} & \textbf{Recall} & \textbf{F1-Score} \\
\midrule
0.30 & 0.50 & 1.00 & 0.67 \\
0.35 & 0.51 & 0.99 & 0.67 \\
0.40 & 0.53 & 0.99 & 0.69 \\
0.45 & 0.56 & 0.95 & 0.71 \\
0.50 & 0.62 & 0.87 & \textbf{0.72} \\
0.55 & 0.68 & 0.71 & 0.70 \\
0.60 & 0.75 & 0.49 & 0.59 \\
0.65 & 0.81 & 0.26 & 0.39 \\
0.70 & 0.96 & 0.11 & 0.20 \\
\bottomrule
\end{tabular}
\end{table}

\subsection{Error Modes}
The keyword baseline's false negatives come overwhelmingly from lexical variation---synonym, acronym, and title differences---which is expected, since that variation was introduced deliberately and the labels ignore wording. The semantic pipeline fails differently: its false positives arise mainly where distinct skill sets are described in similar language, a limitation of surface similarity rather than a wording artifact. Closing that gap would take richer, structured candidate signals, not more aggressive matching.

\subsection{Interpretation}
Within this controlled setting, screening outcomes turn heavily on representation, not only on qualification. Exact matching imposes a bottleneck that rejects more qualified candidates as wording diversifies, which shows how part of matching friction could be algorithmically induced---and, in principle, corrected by better representation. The scope holds: these results demonstrate a mechanism under controlled conditions and do not measure the prevalence or economic magnitude of Artificial Frictional Unemployment, which requires field data.

\section{Discussion and Ethical Considerations}
Any system that intervenes in hiring must be judged on more than accuracy. This section addresses fairness, transparency, privacy, and the risk of ceding too much to automation.

\subsection{Bias and Fairness}
Automated recruitment can reproduce existing inequalities through biased data, design choices, or proxies for protected attributes. Semantic matching reduces vocabulary-driven exclusion but does not remove bias on its own. The proposed design mitigates it in two ways: masking demographic and institutional signals during encoding, so names and pedigree do not shape similarity, and using probabilistic scores rather than hard thresholds, which makes calibration inspectable and supports fairness auditing.

\subsection{Transparency}
Opaque decisions erode trust, particularly when candidates cannot learn why they were rejected. The design retains intermediate representations and similarity rationales, allowing it to surface which skill clusters drove an alignment without exposing proprietary internals. Audit logging and versioned updates let outcomes be traced to specific configurations after the fact.

\subsection{Data Ownership and Privacy}
Employment histories are among the most sensitive personal data, and platforms often centralize and monetize them without meaningful consent. The proposed model keeps ownership with the candidate, who controls disclosure and can revoke access, with encryption enforced throughout. The aim is data minimization by design rather than as an afterthought.

\subsection{Over-Automation}
Hiring depends on judgment that resists full formalization. The framework deliberately declines to make JobOS a decision-maker; it improves signal quality ahead of human review and leaves the decision with people. The recall-oriented tradeoff reported above reinforces this: a high-recall, lower-precision tool is suited to surfacing candidates, not to closing the loop automatically.

\subsection{Policy Implications}
At scale, better matching shortens unemployment spells, improves mobility, and raises participation. Read as policy, the framework suggests that some labor market inefficiency reflects choices embedded in software, which makes modernizing recruitment technology a legitimate lever alongside workforce development. The question is interdisciplinary by nature, spanning economics, computer science, and law, and it connects to emerging regulation of automated employment tools~\cite{nyc2021ll144,euaiact2024}.

\subsection{Limitations}
Several limits bear directly on how the results should be read.

The evaluation uses synthetic resumes, not real hiring data. This buys controlled labeling but caps external validity, and the reported magnitudes do not estimate real-world AFU. Real resumes carry noise---inconsistent formatting, missing information, strategic self-presentation---that the simulation omits. Validating the mechanism on anonymized field data is the central next step.

The keyword pipeline is a stylized stand-in, not a specific vendor's system, so absolute numbers should be read with care. The qualitative failure modes it exhibits---vocabulary mismatch and recall collapse---are, however, documented across independent ATS audits.

The study covers early-stage screening only and says nothing about interview success or job performance. Higher recall at the screening stage does not guarantee better hires; JobOS is proposed as an augmentation layer for exactly this reason.

Finally, semantic tools may shift candidate behavior over time toward strategic phrasing. More resistant to gaming than keyword filters, they are not immune, which is why transparency and periodic recalibration matter.

\section{Conclusion}
This paper took up a visible contradiction: high vacancies and long unemployment holding together in the U.S. labor market. Rather than resting on skills mismatch or macroeconomic friction, we argue that part of the inefficiency comes from the design of automated recruitment---deterministic keyword screening that rejects qualified candidates for how they phrase their experience, not for what they can do.

Placing this within labor market theory, we introduced Artificial Frictional Unemployment as a framework for a digitally induced failure in worker--firm matching. A controlled proof-of-concept simulation illustrated the mechanism, showing how replacing exact matching with semantic matching recovers qualified candidates that keyword screening discards. The gain comes with a real cost in precision, which we report rather than hide, and which points to the appropriate use of the method as an augmentation layer feeding human review. We are explicit that the simulation demonstrates a mechanism and does not measure its economic magnitude.

On that foundation we outlined JobOS, a candidate-side architecture that augments existing pipelines instead of replacing them, combining semantic encoding, verification, and candidate-owned governance. The work does not argue for automated hiring decisions; it argues for improving the signal quality on which human decisions rest, with fairness, transparency, and data ownership built in from the start.

Reframing hiring as semantic translation, and automated recruitment as national labor market infrastructure, opens an agenda across computer science, labor economics, and policy---toward matching that is more efficient, more equitable, and more resilient.


\end{document}